\documentclass[runningheads]{llncs}
\usepackage{hyperref}
\usepackage{graphicx}
\usepackage{subcaption}
\usepackage{xcolor}
\begin{document}
%
\title{Multimodal brain tumor classification}
%
%
\author{Marvin Lerousseau\inst{1,2} \and
Eric Deutsch\inst{1} \and
Nikos Paragios\inst{3}}
\authorrunning{M. Lerousseau et al.}
%
\institute{Paris-Saclay University, Gustave Roussy, Inserm, 94800, Villejuif, France \and
Paris-Saclay University, CentraleSupélec, 91190, Gif-sur-Yvette, France \and
TheraPanacea, 75014, Paris, France}
\maketitle              
\begin{abstract}
Cancer is a complex disease that provides various types of information depending on the scale of observation.
While most tumor diagnostics are performed by observing histopathological slides, radiology images should yield additional knowledge towards the efficacy of cancer diagnostics.
This work investigates a deep learning method combining whole slide images and magnetic resonance images to classify tumors.
In particular, our solution comprises a powerful, generic and modular architecture for whole slide image classification.
Experiments are prospectively conducted on the 2020 Computational Precision Medicine challenge, in a 3-classes unbalanced classification task.
We report cross-validation (resp. validation) balanced-accuracy, kappa and f1 of 0.913, 0.897 and 0.951 (resp. 0.91, 0.90 and 0.94).
For research purposes, including reproducibility and direct performance comparisons, our finale submitted models are usable off-the-shelf in a Docker image available at \url{https://hub.docker.com/repository/docker/marvinler/cpm_2020_marvinler}.

\end{abstract}
\section{Introduction}
Gliomas are the most common malignant primary brain tumor. They start in the glia, which are non-neuronal cells from the central system that provide supportive functions to neurons. There are three types of glia, yielding different types of brain tumor: astrocytomas develop from astrocytes, tumors starting in oligodendrocytes lead to oligodendrogliomas, and ependymomas develop from ependymal cells. The most malignant form of brain tumors is glioblastoma multiforme, which are grade IV astrocytomas. Additionally, some brain tumors may arise from multiple types of glial cells, such mixed gliomas, also called oligo-astrocytomas.

In the clinical setting, the choice of therapies is highly influenced by the tumor grade~\cite{louis20162016}. In such a context, glioblastoma are grade IV, astrocytoma are grade II or III and oligodendroglioma are grave II. 
With modern pushes towards precision medicine, a finer characterization of the disease is considered for treatment strategies. Such therapeutic strategies can be surgery, radiation therapy, chemotherapy, brachytherapy or their combinations such as surgery followed by neoadjuvant radiation therapy. Historically, the classification of brain tumors has relied on histopathological inspection, primarily characterized with light-microscopy observation of H\&E-stained sections, as well as immunohistochemistry testing. This histological predominance has been contrasted with the recent progress in understanding the genetic changes of tumor development of the central nervous system. Additionally, while it is not used in neoplasm diagnostic, neuro-imaging can bring other information regarding the state of the disease.

This work deals with a novel strategy for classifying tumors with several image types. We present a strategy to classify whole slide images and/or radiology images extracted from a common neoplasm. Experiments are conducted on a dataset made of pairs of histopathological images and MRIs, for a 3-categories classification of brain tumor.

\subsection{Related work}
\subsubsection{Whole slide image classification}
Whole slide images (WSI) are often at gigapixel size, which drastically impede their processing with common computer vision strategies. By essence, a WSI is obtained at a certain zoom (called magnification), from which lower magnifications can be interpolated. Therefore, a first strategy for WSI classification would consist in downsampling a full-magnification image sufficiently to be processed by common strategies. However, since the constituent elements of WSIs are the biological cells, downsampling should be limited or the loss of information could completely prevent the feasibility of the task.

One first strategy therefore consists of first classification tiles extracted from a WSI, which is equivalent to WSI segmentation, and then combining those tile predictions into a slide prediction. This was investigated for classifying glioma in~\cite{hou2016patch}, breast cancer in~\cite{gecer2018detection}, lung carcinoma in~\cite{coudray2018classification}, and prostate cancer, basal cell carcinoma, and breast cancer metastases to axillary lymph nodes in~\cite{campanella2019clinical}. More generally, the segmentation part of these works has been unified in~\cite{lerousseau2020weakly}. For combining tile predictions into slide predictions, numerous strategies exist such as ensembling or many other techniques that fall under the umbrella of instance-based multiple instance learning~\cite{dietterich1997solving}.

One limitation with such a decoupled approach is that the tile classifier method does not learn useful features to embed tiles into a latent space which could be sampled by a slide classification method. A generalization consists in having a first model that converts tiles into an embedded (or latent) vector, and a second model which combines multiples such vectors from multiple tiles into a single WSI prediction. This is known as embedded-based multiple instance learning~\cite{dietterich1997solving}. The function that maps the instance space to the slide space can be max-pooling, average-pooling, or more sophisticated functions such as noisy-or~\cite{zhang2006multiple}, noisy-and~\cite{kraus2016classifying}, log-sum-exponential~\cite{ramon2000multi}, or attention-based~\cite{ilse2018attention}.

\subsubsection{Magnetic resonance imaging classification}
There are two major approaches for MRI classification. The first is called radiomics~\cite{lambin2012radiomics} and consists in first extracting a set of features for all images of a training set. Such features typically consist in clinical features such as age, first-order features such as descriptors of tumor shape or volume, and second order features describing textural properties of a neoplasm. These features are then used by common machine learning algorithms as a surrogate of crude MRIs. Examples of such approaches for brain tumor classification are illustrated in~\cite{zhou2018radiomics} and ~\cite{kotrotsou2016radiomics}.
The major limitation of radiomics lies in the fact that tumor volumes must be annotated beforehand in order to extract meaningful features.

The other general approach of tumor classification from MRIs consists in using end-to-end deep learning systems. Deep learning bypasses the necessity of delineating the tumor volume and is more flexible than traditional machine learning since features are learned on-the-fly. Compared to machine learning methods, deep learning is known to require more training samples, but out-competes the former when the number of data is sufficient. The majority of variability of brain tumor classification studies relying on deep learning lies in the architecture used. Examples of such studies are~\cite{talo2019application} or others identified in ~\cite{tandel2019review}.

\subsubsection{Combined radiographic and histologic brain tumor classification}
The 2018 instance of the Computational Precision Medicine Radiology-pathology challenge~\cite{kurc2020segmentation} is the first effort towards providing a publicly available dataset with pairs of WSIs and MRIs. It provided 32 training cases as well as 20 testing cases, balanced between the two brain tumor classes oligodendroglioma and astrocytoma. 3 methods were developed by the participating teams. The top-performing method~\cite{bagari2018combined} used a soft-voting ensemble based on a radiographic model and a histologic model. The MRI model relies on radiomics with prior automatic tumor delineation, while the histologic model classifies tiles extracted from WSIs which priorly are filtered using an outlier detection technique. The second-best performing team~\cite{momeni2018dropout} also used two models for both modalities. The radiographic model is an end-to-end deep learning approach, while the histologic model uses a deep learning model to perform feature extraction by dropping its last layer, to further classify a set of extracted tiles into a WSI prediction. While these two models are trained separately, authors combine features extracted from both of them into an SVM model for case classification, that relies on dropout as regularization to counter the low number of training samples. Finally, the third best performing team used a weighted average of predictions obtained with two end-to-end deep-learning-based classification models for both histologic~\cite{qi2018label} and radiographic modalities, where weights are empirically estimated.


\section{Methods}
Our proposed method leverages both imaging and histological modalities through an ensemble. Specifically, a first deep learning model is intended to classify WSIs, while a second network classifies MRIs.
\subsection{Whole slide image classification with a generic and modular approach}

\begin{figure}[ht]
\includegraphics[width=\textwidth]{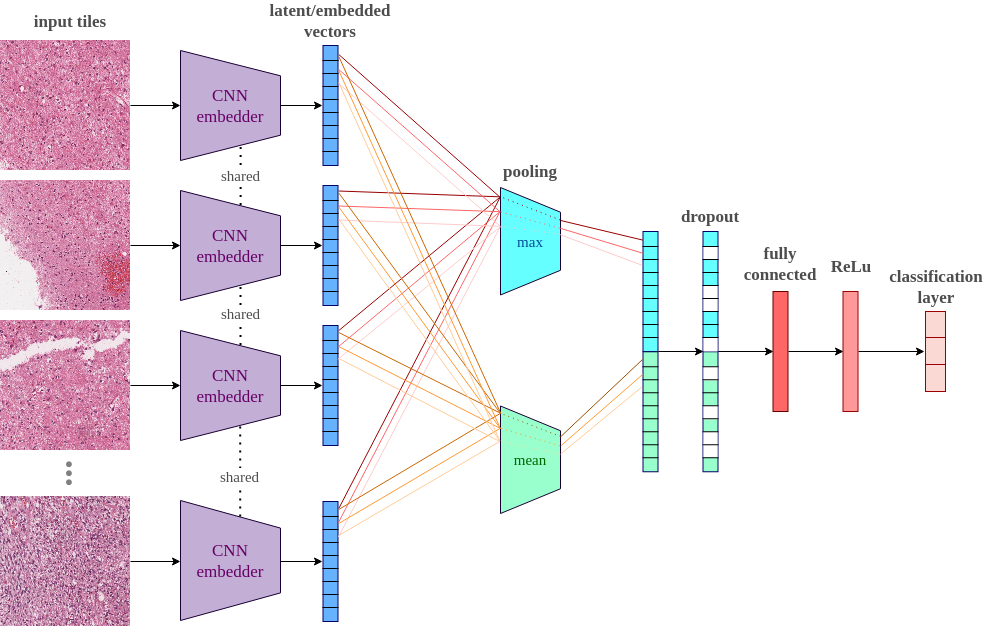}
\caption{Schematic representation of the end-to-end WSI classification strategy. A set of tiles is extracted from a WSI and are each inferred into a unique learnable CNN embedder (i.e. CNN classifier whose last layer has been discarded). Each of the tiles is thus converted into a latent vector (blue vectors) of size $L$, where $L=9$ in this figure. Each latent dimension is then maxed, and simultaneously averaged, into two vectors of size $L$, which are then concatenated into a vector of size $2L$. This WSI latent vector is then forwarded into a dropout and a finale classification layer.} 
\label{fig:histo_archi}
\end{figure}

Our WSI classifier bears most concepts in multiple instance learning approaches as depicted in Fig.\ref{fig:histo_archi}. At training and inference, a WSI is split into non-overlapping tiles. Each tile is forwarded into a standard 2D image classifier, such as a ResNet~\cite{he2016identity} or an EfficientNet~\cite{tan2019efficientnet}, whose classifier layer has been removed. Each tile is thus embedded into a latent vector of size $L$. For a bag of $n$ tiles, a latent matrix of size $n\times L$ is obtained. Then, a max-pooling operation is performed on each feature - that is, across the $n$ dimension. Similarly, average pooling is performed for each feature. Both max-pooled and average-pooled are concatenated, making a resulting vector of size $L+L=2L$. Finally, a classifier network ("head") processes this pooled latent vector into a 1 class prediction. This head is made of a dropout layer, followed by a final classification layer with softmax activation.

This system can be implemented end-to-end in a single network, for standard deep learning optimization. The head part can be more sophisticated, with the addition of an extra linear layer, followed by an activation function such as ReLu, and batch normalization. This implementation was kept simple due to the low number of training samples in our experiments. 
Of major interest, the slide embedding output of size $2L$ is independent of the size $n$ of the input bag. Consequently, bags of any size can be fed into the network during training or inference.

During inference, it is common that the number of non-overlapping tiles is above the memory limit induced by the network, which is roughly the memory footprint of the embedding model. For instance, the number of tiles per slide is depicted in Fig.~\ref{fig:tiled_per_slide} for the training samples at magnification 20. Typically, for EfficientNet\_b0 as an embedding model, only 200 tiles can be fitted in a 16Gb graphic card. Therefore, during inference, a random set of tiles is sampled from a WSI, each yielding one predicted class. The resulting finale predicted class is obtained by hard-voting the latter.

\subsection{Magnetic resonance imaging classification}
\begin{figure}[h]
\includegraphics[width=\textwidth]{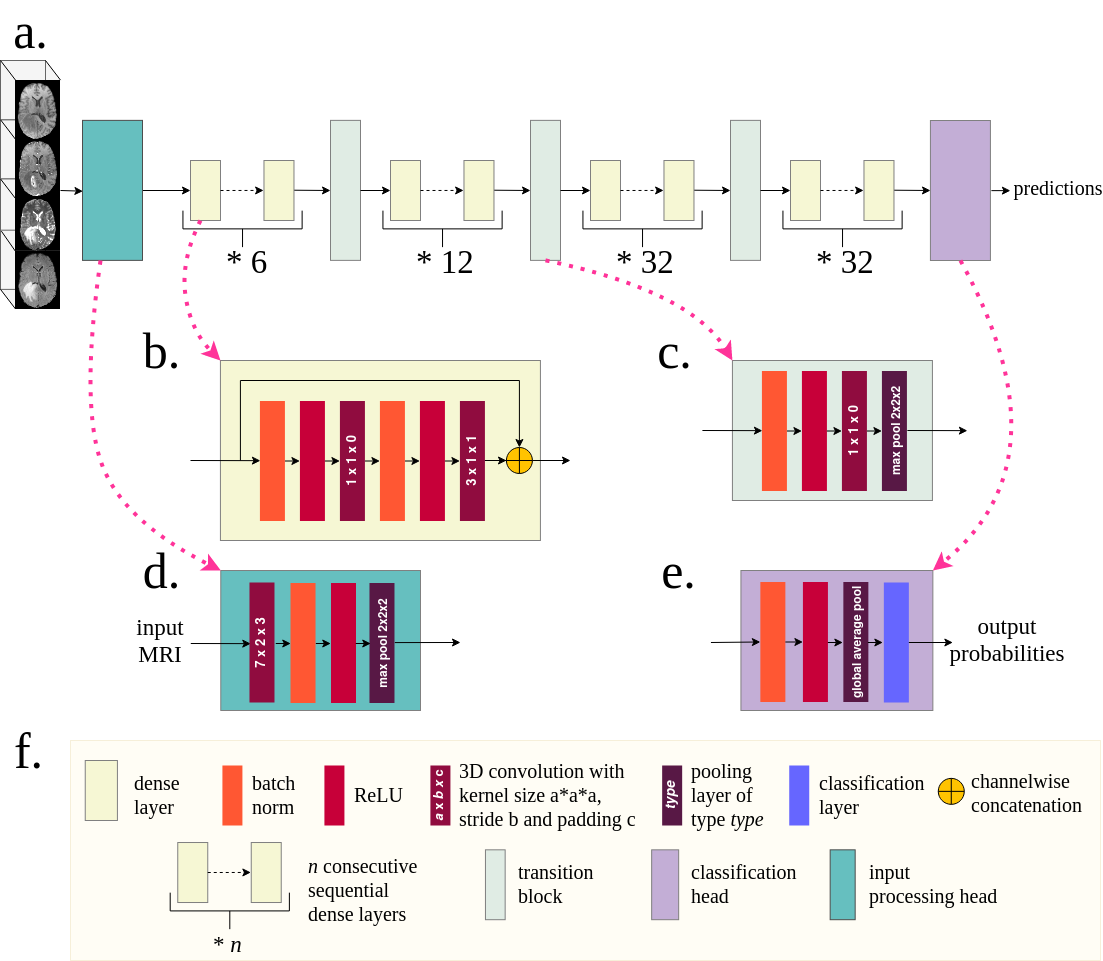}
\caption{Illustration of the Densenet169 architecture made of 169 convolutional layers. The macro representation (a.) shows a feature extractor module made of an input processing head (d.) followed by a combination of dense (b.) and transition (c.) layers. The extracted features are then processed into probabilities by a classification head (e.). All elements are detailed in the legend (f.).} 
\label{fig:densenet}
\end{figure}

Our MRI classification pipeline is straightforward and consists in a single network direct classification of the 4D volumes made of all 4 modalities. Specifically, we use a Densenet~\cite{huang2017densely} made of 169 convolutional layers, whose architecture is displayed in Fig.~\ref{fig:densenet}. The Densenet family was used due to its low number of parameters, which seems appropriate to the low number of experiment training samples, as well as its high number of residual connections which alleviate much gradient issues.
To accommodate with the 3D spatial dimensions of input volumes, the 2D convolutions, and the pooling operators have been modified to 3D. Furthermore, all convolutional kernels are cubic, i.e. with the same size in all 3 dimensions. Various kernel sizes are used throughout the architecture, as specified in Fig.~\ref{fig:densenet}. In particular, inputs MRI made of 4 3-dimensional modalities are stacked on their modality dimension such that the first convolution treats MRI modalities as channels. In practice, the first convolution is of kernel size 7 with stride 2, padding 3 and 64 channels, effectively converting an input MRI volume of size $(4, 128, 128, 128)$ into a output volume of size $(64, 64, 64, 64)$.

\subsection{Multimodal classification through ensembling}
Ensembling was performed for multiple reasons. First of all, without any a priori, histopathological classification could detect different features at various scales, thus producing different or complementary diagnostics depending on the input magnification. In our implementation, multiple networks were trained at various magnifications. Secondly, ensembling allow to use both histopathological and radiological modalities to determine a final diagnostic. Finally, ensembling several neural networks is known to significantly improve the performance and robustness of a neural network system~\cite{hansen1990neural}. 
For these reasons, our final classification uses an ensemble of multiple histopathological networks, radiological networks in a soft-voting way. 
Besides, as shown in~\cite{zhou2002ensembling}, it may be advantageous to ensemble some of the at-hand neural networks rather than all of them. As discussed in \ref{subsection:impl_details}, a portion of the trained networks is discarded in the finale decision system.

\section{Experiments}
Experiments were conducted during the 2020 Computational Precision Medicine Radiology-Pathology challenge (CPM-RadPath 2020). All the data is provided by the organizers, i.e. a training set and an online validation set. Additionally, final results are computed on a hidden testing set with only one try, such as to minimize testing fitting. Our team name was \textit{marvinler}.
\subsection{Data}
\subsubsection{Dataset}
The training data consists in 221 cases extracted from two cohorts: The Cancer Image Archive (TCIA) and the Center for Biomedical Image Computing and Analytics (CBICA). For each case, one formalin-fixed, paraffin-embedded (FFPE) resection whole slide image was provided, along with one MRI, which consists in 4 modalities: T1 (spin-lattice relaxation), T1-contrasted, T2 (spin-spin relaxation), and fluid attenuation inversion recovery (FLAIR) pulse sequences.

\begin{figure}[h]
\includegraphics[width=\textwidth]{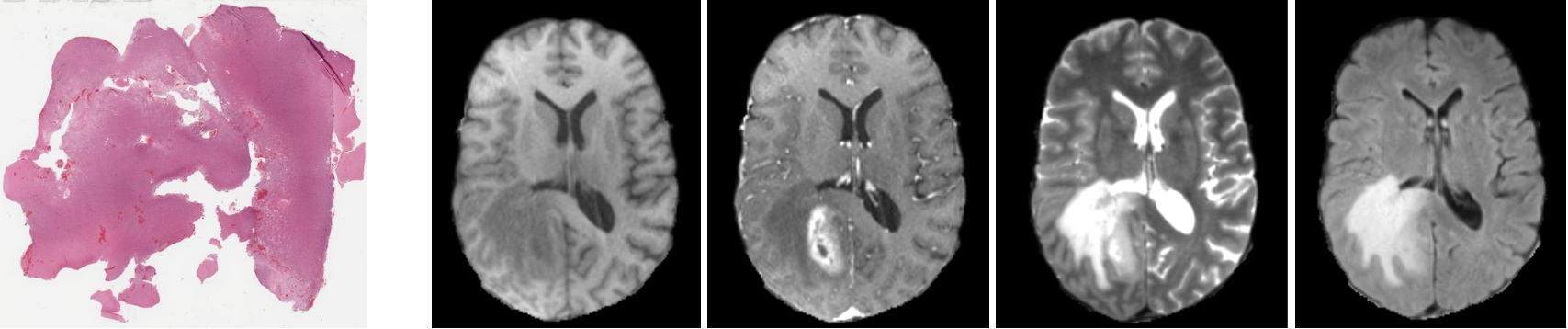}
\caption{Example of data point from the online validation set. The image on the left is a downscaled version of the whole slide image which is initially of width 96224 pixels and of height 82459 pixels. The four images on the right represent the MRI of the same case, and are, by order, the T1, the T1 contrast-enhanced, the T2, and the FLAIR modalities. For the MRI, the same slice extracted from each 3D volume is represented. Each 3D volume is initially of size 240x240x155 pixels.} 
\label{fig:example_data}
\end{figure}

Each case belongs to one of three diagnostic categories among astrocytomas, oligodendrogliomas, and glioblastoma multiforme. This information was provided for each training case. On top of that, a similar validation set made of 35 cases was available for generalization assessment. For these cases, ground-truth labels were hidden, although up to 50 submissions were possible, with feedback containing balanced accuracy, f1 score, and kappa score. Fig.~\ref{fig:example_data} shows an example from the validation set.

\subsubsection{Pre-processing}
No data pre-processing was performed on MRIs which have already been skull-stripped by data providers. Further MRI processing was subcontracted to the radiology data augmentation step, as detailed in \ref{subsection:impl_details}.

\begin{figure}[h]
\includegraphics[width=\textwidth]{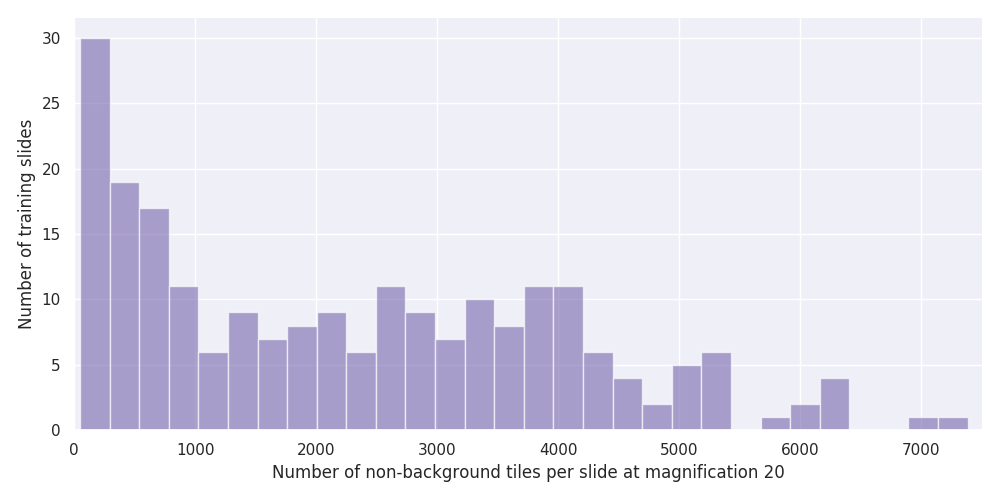}
\caption{Histogram of the number of tiles considered non-background for all training slides at magnification 20, or equivalently micrometer per pixel (mpp) of 0.5.} 
\label{fig:tiled_per_slide}
\end{figure}

Whole slide images were available in non-pyramidal .tiff format. They were first tiled in a pyramidal scheme using libvips, with a tile width of 512 pixel and no tile overlap. For each resulting magnification level, all tiles considered background were discarded. This was done by detecting tiles where at least 75\% of pixels have both red, green, and blue channels above a value of 180 (where 255 is absolute white and 0 is black), which saved more than two third of disk space by discarding non informative tiles from further processing. After this filtering, the number of tiles per slide is depicted in the histogram of Fig.~\ref{fig:tiled_per_slide} for magnification 20. The complete WSI pre-processing was applied to both training, validation, and testing sets.

\subsection{Implementation details}
\label{subsection:impl_details}
A pre-trained Efficientnet\_b0 was selected as the histopathological model embedder, converting 224 pixel-wide images into an embedded vector of size $L=1280$. During training, 50 tiles were selected from 4 WSIs, resulting in a batch of size 200. Each tile was data augmented with random crop from 512 pixel width to 224 pixel width, color jitter with brightness, contrast, saturation of $0.1$, and hue of $0.01$, and was normalized by dividing the mean and standard deviation of each RGB channel as computed on the training dataset. To counter the low number of training samples, dropout of $0.5$ was used in the head of the network. The final softmax activation was discarded during training, and cross-entropy loss was used to compute the error signal (which contains a log softmax operation for numerical stability). To handle class imbalance, weights were computed as the inverse of the frequency of each of the 3 classes and used in the error computation. Adam~\cite{kingma2014adam} optimizer was used to back-propagate the error signal, with default momentum parameters and learning rate of $5e-5$ for 50 epochs. This process was applied for the 5 magnifications of 40, 20, 10, 5, and 2.5, or equivalently and respectively of micrometers per pixel of 0.25, 0.5, 1, 2, and 4. During inference, 200 tiles were randomly sampled from each WSI and for all magnifications, with data augmentation independently applied from tile to tile. Due to poor results on the online validation set, the models from magnification 40 were discarded.

For the classification from MRIs, data augmentation first consisted of cropping the foreground, which roughly crops the MRI to their contained brain. These cropped volumes were resized to a unique size of $128\times128\times128$, and each modality was standard scaled such that 0 values (corresponding to background) were not computed in mean and standard deviation computations. Then, a random zoom between $0.8$ and $1.2$ was applied, followed by a random rotation of $10$ degrees on both sides for all dimensions. Random elastic deformations with parameter values of sigma in range $(1, 10)$ and magnitude in range $(10, 200)$. All of the data augmentation was implemented using the Medical Open Network for AI (MONAI) toolkit\footnote{https://github.com/Project-MONAI/MONAI}, which is also the source of the Densenet169 architecture implementation. Error signal was computed with cross-entropy loss and back-propagated with Adam optimizer for 200 epochs at learning rate $5e-4$ with a batch size of $3$. The same class imbalance was used than the histopathological networks.

For both 5 models (4 histopathological and 1 radiological), rather than selecting the best performing models on cross-validation, two snapshots in a well performing region were selected. Specifically, the last epoch weights snapshot, as well as the snapshot from the 10 epochs to the end were collected for each model. Following~\cite{zhou2002ensembling}, the 2 least performing (on the local validation) out of the 10 networks were discarding, which consisted in one radiology-based model and one model from magnification 5. Soft-voting was performed on the 8 resulting models for all slides, and the class with highest probability was assigned as the predicted class. 

\subsection{Results}
After 3 online submissions, the hidden validation performance was a balanced accuracy of 0.911, a kappa score of 0.904, and an f1 score (with micro average) of 0.943 which ranked us second overall. Unseen testing results are still awaiting. Our cross-validation results are similar, with a balanced accuracy of 0.913, a kappa score of 0.897 and an f1 score of 0.951, denoting a certain high generalization capacity from our approach.

\begin{figure}[h]
     \centering
     \begin{subfigure}[b]{0.72\textwidth}
        \centering
        \includegraphics[width=\textwidth]{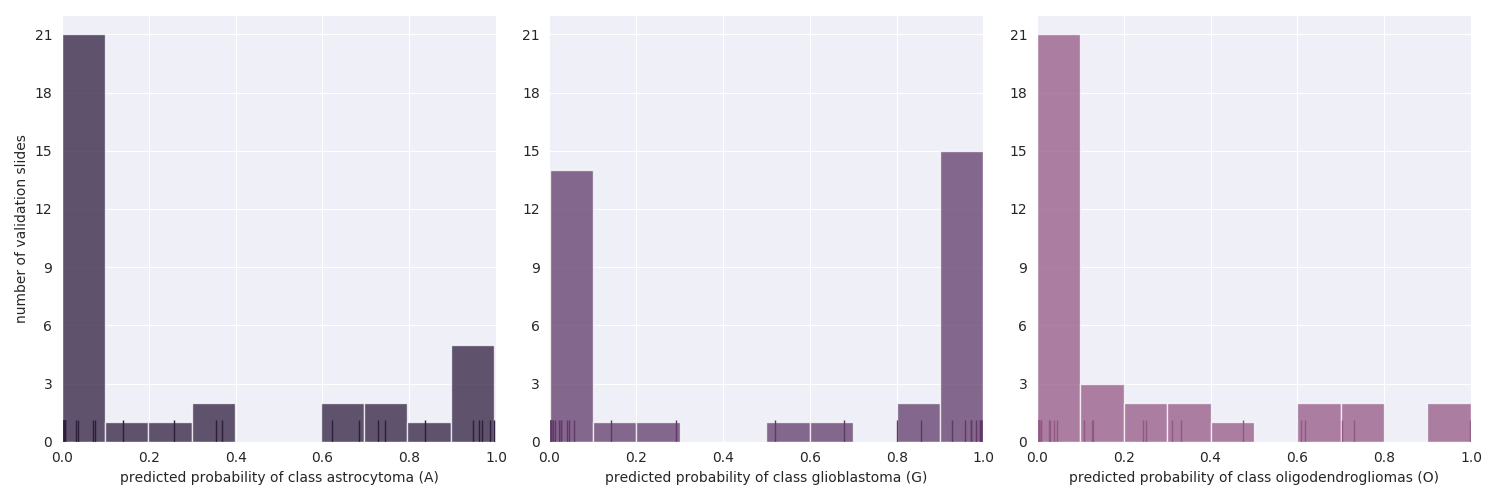}
        \caption{}
        \label{fig:consensusa}
     \end{subfigure}
     \hfill
     \begin{subfigure}[b]{0.24\textwidth}
        \centering
        \includegraphics[width=\textwidth]{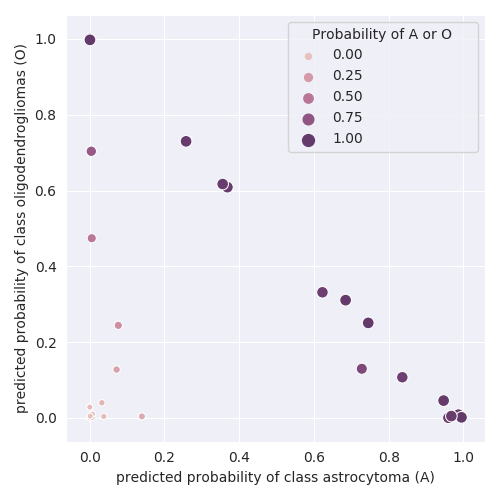}
        \caption{}
        \label{fig:consensusb}
     \end{subfigure}
    \caption{Analysis of validation predicted probabilities. (a) Histogram of predicted probabilities for all 35 validation cases. Each plot (3 in total) refers to one of the 3 predicted classes, as indicated in the x-axis label. (b) Scatterplot of predictions of astrocytoma vs oligodendrogliomas. Color and size of points are proportional to the sum of predictions of the latter class, or equivalently inversely proportional to the predicted probability of glioblastoma multiforme.} 
\end{figure}

For each diagnostic category, a histogram of validation predicted probabilities was computed, resulting in 3 plots depicted in Fig.~\ref{fig:consensusa}. Notably, there seems to be less hesitation for the prediction of the glioblastoma multiforme class (G), where only 3 cases were predicted with a probability between $0.2$ and $0.8$. In comparison, 7 (resp. 9) cases are predicted with a probability between $0.2$ and $0.8$ for the astrocytoma (A) (resp. oligodendrogliomas (O)). Fig.~\ref{fig:consensusb} further highlights that most hesitation comes from both classes A and O, with all probabilities of class A that are not below $0.2$ or above $0.8$ predicted mostly as class O compared to G. Besides, $6$ out of $9$ unsure probabilities for the O class were split with the A class compared to the G one. This could illustrate that some cases seem to exhibit both oligodendrogliomas and astrocytoma, known as mixed gliomas. Notably, while mixed gliomas are a valid diagnostic in clinical settings, this class was absent in the challenge data, with a decision between class A and O taken by the challenge annotators for such cases.

An ablation study was performed on the online validation set to understand the impact of both radiographic and histologic models. The MRI-only model obtained a balanced accuracy of 0.694, a kappa score of 0.662 and an f1 score of 0.800, which is significantly lower than the WSI-only model which had the same performance than our finale model combining both MRI and WSI modalities. However, upon inspection of class predictions, there were 2 differences between the predictions of the latter and the ensemble of WSI-only models. Specifically, one ensemble predicted an oligodendroglioma while the other predicted astrocytoma, and the opposite. Both WSI-only and WSI coupled with MRI models had a consensus on the glioblastoma class. 


\section{Conclusion \& Discussion}
Our work illustrates the feasibility of the classification of tumor types among 3 pre-defined categories. Although testing results are still pending, the proposed approach appears to generalize well to unseen data from the same distribution. The proposed pipeline heavily relies on histopathological slides, which have been the golden standard for tumor diagnosis for a hundred years. This system is an end-to-end trainable decision system that relies on a learnable embedding model (e.g. a convolutional neural network) and combines a set of any size of tiles embedded vectors into a slide latent vector for further classification. This generic deep learning pipeline can be taken off-the-shelf and applied to many other histopathological classification tasks, would it be grading, diagnostic, primary determination or prognosis purposes. For this reason, we open-source our complete whole slide image classification method in a Docker image at \url{https://hub.docker.com/repository/docker/marvinler/cpm_2020_marvinler}, ensuring its use off-the-shelf on any platform. Such a resource can be used to perform direct comparisons of future research contributions on combined WSI and MRI brain tumor classification, while also ensuring reproducibility of results. 

While our use of the radiology modality has not been heavily in this work, there should be improvement in classifying cases with more information than in a histopathological context only. We believe that, with more training data, diagnosis could be further improved with multimodal solutions that embed both radiographic images and histologic images into a common latent space.

%
%
%
\bibliographystyle{splncs04}
\bibliography{biblio}

\begin{thebibliography}{10}
\providecommand{\url}[1]{\texttt{#1}}
\providecommand{\urlprefix}{URL }
\providecommand{\doi}[1]{https://doi.org/#1}

\bibitem{bagari2018combined}
Bagari, A., Kumar, A., Kori, A., Khened, M., Krishnamurthi, G.: A combined
  radio-histological approach for classification of low grade gliomas. In:
  International MICCAI Brainlesion Workshop. pp. 416--427. Springer (2018)

\bibitem{campanella2019clinical}
Campanella, G., Hanna, M.G., Geneslaw, L., Miraflor, A., Silva, V.W.K., Busam,
  K.J., Brogi, E., Reuter, V.E., Klimstra, D.S., Fuchs, T.J.: Clinical-grade
  computational pathology using weakly supervised deep learning on whole slide
  images. Nature medicine  \textbf{25}(8),  1301--1309 (2019)

\bibitem{coudray2018classification}
Coudray, N., Ocampo, P.S., Sakellaropoulos, T., Narula, N., Snuderl, M.,
  Feny{\"o}, D., Moreira, A.L., Razavian, N., Tsirigos, A.: Classification and
  mutation prediction from non--small cell lung cancer histopathology images
  using deep learning. Nature medicine  \textbf{24}(10),  1559--1567 (2018)

\bibitem{dietterich1997solving}
Dietterich, T.G., Lathrop, R.H., Lozano-P{\'e}rez, T.: Solving the multiple
  instance problem with axis-parallel rectangles. Artificial intelligence
  \textbf{89}(1-2),  31--71 (1997)

\bibitem{gecer2018detection}
Gecer, B., Aksoy, S., Mercan, E., Shapiro, L.G., Weaver, D.L., Elmore, J.G.:
  Detection and classification of cancer in whole slide breast histopathology
  images using deep convolutional networks. Pattern recognition  \textbf{84},
  345--356 (2018)

\bibitem{hansen1990neural}
Hansen, L.K., Salamon, P.: Neural network ensembles. IEEE transactions on
  pattern analysis and machine intelligence  \textbf{12}(10),  993--1001 (1990)

\bibitem{he2016identity}
He, K., Zhang, X., Ren, S., Sun, J.: Identity mappings in deep residual
  networks. In: European conference on computer vision. pp. 630--645. Springer
  (2016)

\bibitem{hou2016patch}
Hou, L., Samaras, D., Kurc, T.M., Gao, Y., Davis, J.E., Saltz, J.H.:
  Patch-based convolutional neural network for whole slide tissue image
  classification. In: Proceedings of the ieee conference on computer vision and
  pattern recognition. pp. 2424--2433 (2016)

\bibitem{huang2017densely}
Huang, G., Liu, Z., Van Der~Maaten, L., Weinberger, K.Q.: Densely connected
  convolutional networks. In: Proceedings of the IEEE conference on computer
  vision and pattern recognition. pp. 4700--4708 (2017)

\bibitem{ilse2018attention}
Ilse, M., Tomczak, J.M., Welling, M.: Attention-based deep multiple instance
  learning. arXiv preprint arXiv:1802.04712  (2018)

\bibitem{kingma2014adam}
Kingma, D.P., Ba, J.: Adam: A method for stochastic optimization. arXiv
  preprint arXiv:1412.6980  (2014)

\bibitem{kotrotsou2016radiomics}
Kotrotsou, A., Zinn, P.O., Colen, R.R.: Radiomics in brain tumors: an emerging
  technique for characterization of tumor environment. Magnetic Resonance
  Imaging Clinics  \textbf{24}(4),  719--729 (2016)

\bibitem{kraus2016classifying}
Kraus, O.Z., Ba, J.L., Frey, B.J.: Classifying and segmenting microscopy images
  with deep multiple instance learning. Bioinformatics  \textbf{32}(12),
  i52--i59 (2016)

\bibitem{kurc2020segmentation}
Kurc, T., Bakas, S., Ren, X., Bagari, A., Momeni, A., Huang, Y., Zhang, L.,
  Kumar, A., Thibault, M., Qi, Q., et~al.: Segmentation and classification in
  digital pathology for glioma research: Challenges and deep learning
  approaches. Frontiers in Neuroscience  \textbf{14} (2020)

\bibitem{lambin2012radiomics}
Lambin, P., Rios-Velazquez, E., Leijenaar, R., Carvalho, S., Van~Stiphout,
  R.G., Granton, P., Zegers, C.M., Gillies, R., Boellard, R., Dekker, A.,
  et~al.: Radiomics: extracting more information from medical images using
  advanced feature analysis. European journal of cancer  \textbf{48}(4),
  441--446 (2012)

\bibitem{lerousseau2020weakly}
Lerousseau, M., Vakalopoulou, M., Classe, M., Adam, J., Battistella, E.,
  Carr{\'e}, A., Estienne, T., Henry, T., Deutsch, E., Paragios, N.: Weakly
  supervised multiple instance learning histopathological tumor segmentation.
  arXiv preprint arXiv:2004.05024  (2020)

\bibitem{louis20162016}
Louis, D.N., Perry, A., Reifenberger, G., Von~Deimling, A., Figarella-Branger,
  D., Cavenee, W.K., Ohgaki, H., Wiestler, O.D., Kleihues, P., Ellison, D.W.:
  The 2016 world health organization classification of tumors of the central
  nervous system: a summary. Acta neuropathologica  \textbf{131}(6),  803--820
  (2016)

\bibitem{momeni2018dropout}
Momeni, A., Thibault, M., Gevaert, O.: Dropout-enabled ensemble learning for
  multi-scale biomedical data. In: International MICCAI Brainlesion Workshop.
  pp. 407--415. Springer (2018)

\bibitem{qi2018label}
Qi, Q., Li, Y., Wang, J., Zheng, H., Huang, Y., Ding, X., Rohde, G.K.:
  Label-efficient breast cancer histopathological image classification. IEEE
  journal of biomedical and health informatics  \textbf{23}(5),  2108--2116
  (2018)

\bibitem{ramon2000multi}
Ramon, J., De~Raedt, L.: Multi instance neural networks. In: Proceedings of the
  ICML-2000 workshop on attribute-value and relational learning. pp. 53--60
  (2000)

\bibitem{talo2019application}
Talo, M., Baloglu, U.B., Y{\i}ld{\i}r{\i}m, {\"O}., Acharya, U.R.: Application
  of deep transfer learning for automated brain abnormality classification
  using mr images. Cognitive Systems Research  \textbf{54},  176--188 (2019)

\bibitem{tan2019efficientnet}
Tan, M., Le, Q.V.: Efficientnet: Rethinking model scaling for convolutional
  neural networks. arXiv preprint arXiv:1905.11946  (2019)

\bibitem{tandel2019review}
Tandel, G.S., Biswas, M., Kakde, O.G., Tiwari, A., Suri, H.S., Turk, M., Laird,
  J.R., Asare, C.K., Ankrah, A.A., Khanna, N., et~al.: A review on a deep
  learning perspective in brain cancer classification. Cancers  \textbf{11}(1),
  ~111 (2019)

\bibitem{zhang2006multiple}
Zhang, C., Platt, J.C., Viola, P.A.: Multiple instance boosting for object
  detection. In: Advances in neural information processing systems. pp.
  1417--1424 (2006)

\bibitem{zhou2018radiomics}
Zhou, M., Scott, J., Chaudhury, B., Hall, L., Goldgof, D., Yeom, K.W., Iv, M.,
  Ou, Y., Kalpathy-Cramer, J., Napel, S., et~al.: Radiomics in brain tumor:
  image assessment, quantitative feature descriptors, and machine-learning
  approaches. American Journal of Neuroradiology  \textbf{39}(2),  208--216
  (2018)

\bibitem{zhou2002ensembling}
Zhou, Z.H., Wu, J., Tang, W.: Ensembling neural networks: many could be better
  than all. Artificial intelligence  \textbf{137}(1-2),  239--263 (2002)

\end{thebibliography}
\end{document}